# Discovery of voltage induced superfluid-like penetration effect in liquid metals at room temperature

**Frank F. Yun[1], Zhenwei Yu[1], Yahua He[1], Lei Jiang[2], Haoshuang Gu[3], Zhao Wang[3], Xiaolin Wang[1, 4]∗**

[1]Institute for Superconducting and Electronic Materials, Australian Institute for Innovative Materials, University of Wollongong, Wollongong, New South Wales 2500, Australia.

[2]Laboratory of Bioinspired Smart Interfacial Science, Technical Institute of Physics and Chemistry, Chinese Academy of Sciences, Beijing 100190, People's Republic of China.

[3]Faculty of Physics & Electronic Sciences, Hubei University, Wuhan, 430062 P.R. China

[4]ARC Centre of Excellence in Future Low-Energy Electronics Technologies, University of Wollongong, Wollongong, New South Wales 2500, Australia.

**Frank F. Yun and Zhenwei Yu contributed equally to this work.**

**Abstract**

We have discovered that room temperature liquid metal is capable of penetrating through macro- and microporous materials by applying a voltage. In this work, we demonstrate the liquid metal penetration effect in various porous materials such as tissue paper, thick and fine sponges, fabrics, and meshes. The penetration effect mimics one of the three well-known superfluid properties of liquid helium superfluid that only occur at near-zero Kelvin. The underlying mechanism is that the high surface tension of liquid metal can be significantly reduced to near-zero due to the voltage induced oxidation of the liquid metal surface in a solution. It is the extremely low surface tension and gravity that cause the liquid metal to superwet the solid surface, leading to the penetration phenomena. Our findings offer new opportunities for novel microfluidic applications and could promote further discovery of more exotic fluid states of liquid metals.

**Main Text**

Penetration through a solid with micro- or nanopores is one of the three fascinating macroscopic phenomena that are well known in superfluids such as liquid helium[1-3]. It is the zero viscosity that endows the liquid helium superfluid with the ability to flow without any resistance, leading to its amazing penetration and other superfluid effects. Nevertheless, the helium superfluid penetration effect emerges only in the so-called quantum states or quantum fluids at extremely low temperatures of almost zero Kelvin. In contrast, conventional liquids such as water and oils can diffuse into or penetrate through a solid with macro-pores at room temperature as a result of the capillary effect, although their surface tensions are not low enough to enable them to penetrate through porous materials. In this work, we propose to study, for the first time, possible penetration effects of gallium based liquid metals, since their surface tension can be significantly and easily tuned by voltage at room temperature.

Recently, room-temperature liquid metals, such as liquid gallium and its eutectic alloys, have attracted great attention in many research fields due to their unique chemical and physical properties, such as negligible vapor pressure[4], large surface tension[5], low toxicity[6], and high electrical[7] and thermal[4] conductivity. We highlight some typical phenomena observed in liquid gallium and its alloys under the application of a voltage at room temperature: 1) giant deformation in acid or base solutions[8-10]; 2) self-rotation[11, 12]; 3) locomotion[12, 13]; 4) spontaneous fast deformation and solidification in the supercooled state[14]; 5) the electro-hydrodynamic shooting phenomenon [15], 6) non-contact and maskless electrochemical patterning or lithography [16], 7) the phagocytosis effect [17], 8) the triggered wire oscillation effect[18], and 9) the liquid metal heartbeat effect[19]. Note that phenomena 4, 6, and 9 were recently discovered by our group. It is noteworthy that the surface tension of liquid gallium and its alloys is quite high (~500 mJ/m$^2$)[4], and therefore, they do not wet most solids. When a layer of oxide, e.g., $Ga_2O_3$, is formed on the surface, however, due to the electrochemical reactions in acid or alkaline solutions, the surface tension is then reduced greatly and can reach a near-zero value under an applied-voltage[19]. It is well accepted that the extremely low surface tension plays a key role in the giant deformation effect in liquid metals[9]. Inspired by the near-zero surface tension phenomenon, here, we propose to explore liquid metal's new capability of penetrating through porous materials with the help of both voltage and gravity. In this communication, we demonstrate the liquid metal penetration effect in various porous materials such as tissue paper, fine sponges, fabrics, and meshes.

The experimental set up for the penetration effect is shown in Fig. 1a & b. A Galinstan droplet is placed on the top surface of a porous material which is fixed inside a Teflon container. The whole set-up is immersed in NaOH electrolyte. A copper wire is inserted into the Galinstan droplet to act as an anode, and another copper wire to act as a cathode is placed in the electrolyte. Different porous materials with various average pore sizes and thicknesses were used All experiments were conducted using different voltages and solution concentrations.

We first demonstrate our observations on the liquid metal penetration effect using a plastic mesh 0.2 mm in thickness and with 0.75 mm pore size. The experiments were conducted under a DC voltage of 5 V in 1 mol/L NaOH solution. A 150 μL Galinstan droplet is placed on the top surface of the plastic mesh. The Galinstan droplet remains stationary and has a spheroidal shape. When a voltage is applied, the droplet deforms rapidly within 0.03 s and becomes totally flat at time, $t = 1.4$ s on the plastic mesh surface. When $t > 1.8$ s, it starts to penetrate through the pores of the plastic mesh, and then the droplet continues to flow down, forming thin threads and eventually touching the bottom of the container. Some of the threads are cut off before reaching the bottom, depending on the sizes of the droplets. Snapshots of the penetration effect for the plastic mesh are shown in Fig. 1c.

There are many other porous materials such as fabric meshes, metal meshes, and tissue paper with different pore sizes. We have further verified the liquid metal penetration effect for all these materials. These experiments were conducted under a voltage of 5 V, in a 1 mol/L NaOH solution with all these materials 0.2 mm in thickness (Fig. 2a-d). The Galinstan droplets can penetrate through all these porous materials and flow out from the other side. Snapshots of Galinstan penetrating through a metallic mesh with a pore size of ~45 µm are shown in Fig. 2e. A few typical snapshots showing the penetration effect for wiper paper (with average pore size of a few microns), and fabrics (~280 µm) and plastic (750 µm) meshes are provided in the Extended Data Fig. S1. We conclude that the penetration process takes place for all thin porous materials.

We now further demonstrate the liquid metal penetration phenomenon using thicker sponges with thicknesses of up to 10 mm and different pore sizes. Three sponges with average pore sizes of ~150 (sponge A), ~ 350 (sponge B), and ~550 µm (sponge C) were used to further reveal the penetration effect. Our results show that the larger the pore sizes, the quicker it is for the liquid metal to penetrate through the sponges with the same magnitude of applied DC voltage and the same solution concentration.

Fig. 2f shows snapshots of the penetration effect for a 7.5 mm thick sponge. Remarkably, upon application of a voltage, the droplet immediately merges into the sponge, quickly ($t = 2$ s) leaks out from the other side of the sponge, and continuously flows down to the bottom of the container. This process continues if the voltage is kept on, and it can be stopped quickly, either by turning off the applied voltage or running out of the Galinstan droplet.

To investigate the voltage and electrolyte concentration effects on the liquid metal penetration phenomenon, DC voltages of 2.5 up to 15 V in NaOH solutions with a concentration of 0.25, 0.5, and 1 mol/L were used for sponges A, B, and C, as shown in Fig. 3 a-c. The volume of Galinstan used was fixed at 150 µL for all experiments.

Fig. 3d shows the experimental conditions, which help us to find out under what conditions the liquid metal can penetrate a sponge, such as the pore size of the sponge, the applied voltage, and the concentration of the electrolyte solution. Fig. 3d contains regions with three typical sets of conditions. For green-region conditions, the liquid metals are able to penetrate through all the sponges. For high voltages and low concentrations (red region), however, no penetration is observed, which is likely due to the rapid reaction that produces excessive oxide on the surface of the droplet, which cannot be dissolved in NaOH solution in a timely manner. Moreover, for conditions in the un-shaded area, penetration can only take place for sponges B and C. We can conclude that the penetration process mainly depends on the oxidation/dissolution rate and the pore size. According to our observations, the maximum volume of the Galinstan that can penetrate through a sponge is unlimited under the right conditions, so long as the Galinstan is supplied continuously. We also found that the penetration speed increases with increasing pore size of the sponge for all conditions. For sponge A, the penetration speeds were 0.1 mm/s for 2.5 V in the 0.25 mol/L NaOH solution and 0.34 mm/s for 10 V in the 1 mol/L NaOH solution (Supporting Information (SI) Video 1). The penetration speeds for sponges B and C are in the range of 0.3 - 1.9 and 2.5 - 7.5 mm/s (SI Videos 2 & 3), respectively (Fig. 3e). A few typical snapshots showing the penetration effect under different conditions for different sponges are given in the Extended Data Figs. 2 & 3.

Furthermore, we have monitored the penetration process for all samples with and without cutting off the applied voltage. We found that once the voltage is turned off, the metal droplet stops penetrating and pops back up to the top surface of the porous material due to the recovery of the large surface tension of Galinstan. Simultaneously, the thin metal threads extending down into the solution break up very quickly and re-shape into spherical droplets.

We now discuss the mechanism for the observed fascinating penetration phenomena of the liquid metal for all these porous materials. It is well known that the flow of any liquid is determined by two main factors: the fluid's viscosity and its surface tension. For zero viscosity, the near-zero contact angle, causing the liquid metal droplet to spread very easily on the surface of any solid interface (Fig. 4a and b). This process enables continuous spreading of the liquid metal when in contact with a solid surface, regardless of the size of the contact area. As a result, the liquid metal can easily penetrate into any porous material (Fig. 4c and d).

Galinstan has a high density (~6.359 g/cm$^3$), and hence, gravity can also contribute to its flow. We assume a solid material with an appropriate porosity in which all voids are interconnected. When a droplet of liquid metal is placed on its surface, the liquid metal remains stationary on the surface and will not penetrate through the solid due to its large surface tension. When the surface tension is significantly reduced to near-zero, however, the droplet will start to diffuse into the solid and continues to flow into the interconnected voids until it finally penetrates through the solid, leaks out from other side, and flows down under the force of gravity (Fig. 4d).

There are several potential applications of the liquid metal penetrating effect. Two possible applications are to heal and cut-off electrical wiring using liquid metal in a sealed environment. Here, we demonstrate our experiments for the two possible applications. For the healing effect, the disconnected copper wires are initially placed at the bottom of a container (Fig. 5a), while a light emitting diode (LED) connected to the circuit is off. We then place a liquid Galinstan droplet on the top surface of a sponge which is fixed 7 centimeters above the open circuit, as shown in Fig. 5a. After applying a voltage, the Galinstan droplet penetrates through the sponge and flows down to the bottom. As a result, the two copper wires are physically connected by the liquid Galinstan. The open circuit is then repaired, and the LED light is turned on (Fig. 5b).

We now demonstrate the cut-off effect. A closed electrical circuit is set up using an aluminium wire which is placed at the container bottom and the LED light is on. After the liquid metal penetrates through the sponge, the droplet drips down to the bottom and comes into direct contact with the aluminium wire while the LED light is still on. After about 2 min, however, the aluminium wire is cut by the liquid Galinstan droplet due to the strong chemical reaction with gallium, and then the LED light turns off (Fig. 5c-e).

In summary, we have discovered that liquid metal is capable of penetrating through any porous material at room temperature. We have demonstrated the penetration effect in sponges with different pore sizes and thicknesses as well as other porous materials such as tissue paper and different meshes. The key physics is that there is a giant reduction of the surface tension of the liquid metal to near-zero, induced by the applied voltage. The near-zero surface tension guarantees a contact angle of zero between the liquid metal and the solid surface. This drives the liquid metal to diffuse into the porous material, keep moving through the material, and drip down by gravity. Our findings offer new opportunities for novel microfluidic applications and could promote further exploration for more exotic fluidic states of liquid metals.

**Figures/Figure. legends**

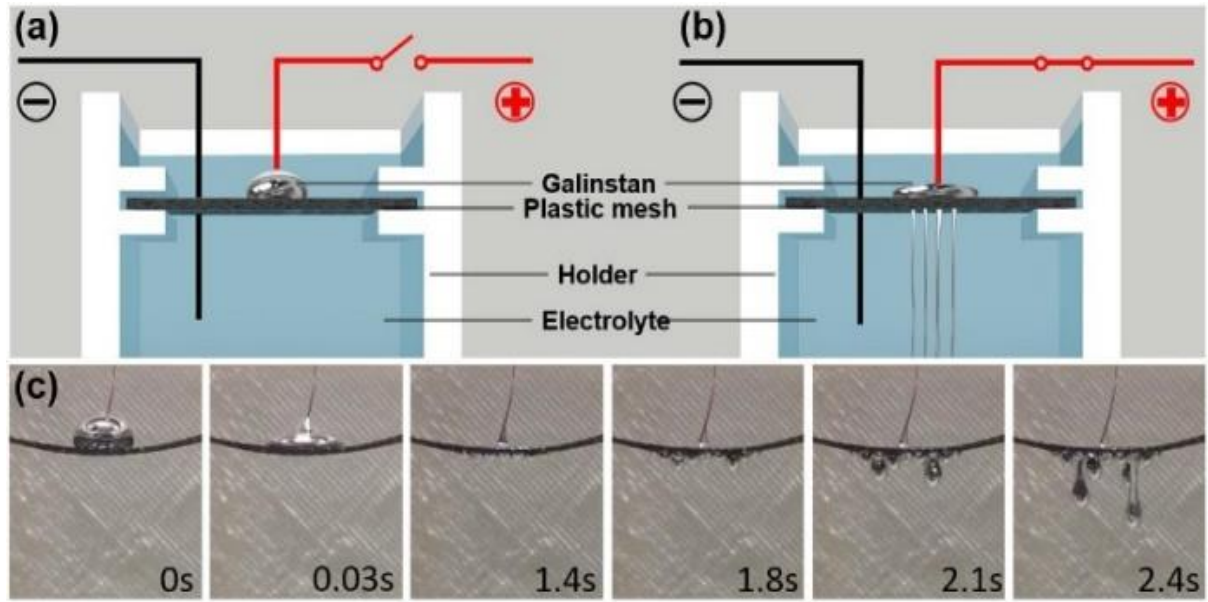

**Fig. 1 | The penetration effect of Galinstan through voltage control in an electrolyte.** Schematic diagram of a galinstan droplet on a plastic mesh before (a) and after (b) the voltage is applied. (c) Snapshots of the penetration effect for a plastic mesh with 5 V applied voltage in 1 mol/L NaOH solution.

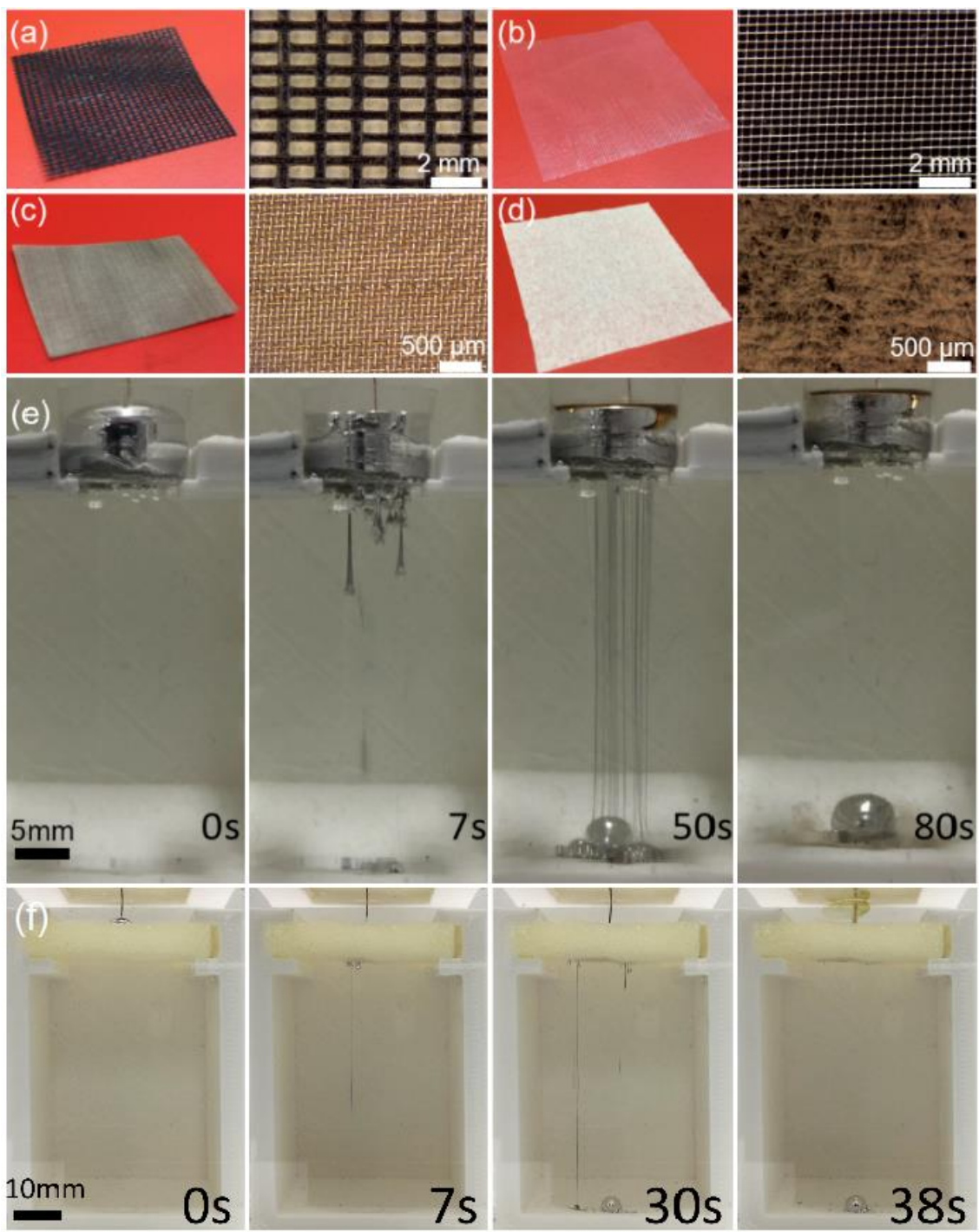

**Fig. 2 | Photographs and optical microscope images** of (a) plastic mesh, (b) fiber mesh, (c) metal mesh, and (d) wiper paper. Snapshots of the penetration effect for (e) metallic mesh with pore size of 45 μm under 5 V applied voltage in 1 mol/L NaOH solution and (f) 7.5 mm thick fine sponge with a 10 V applied voltage in 0.5 mol/L NaOH solution

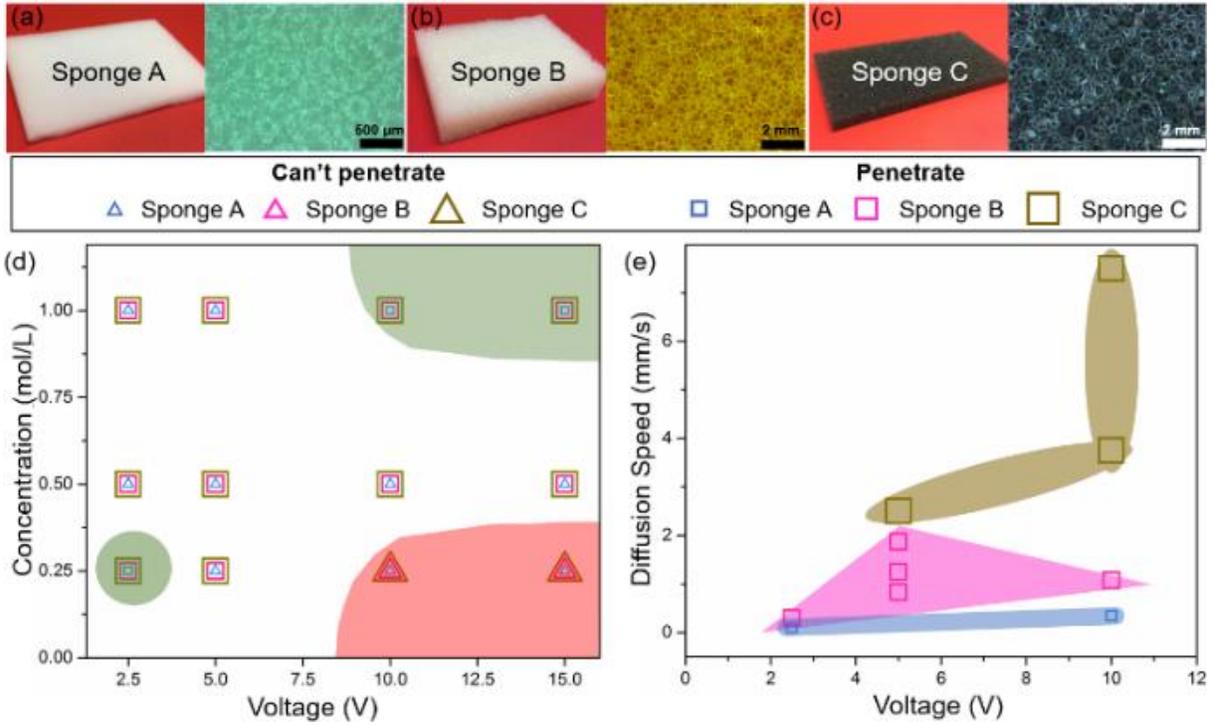

**Fig. 3 | Photographs and optical microscope images** of (a) sponge A, (b) sponge B, and (c) sponge C. (d) Reference diagram of liquid metal penetration effect under different voltages and concentrations of electrolyte. Squares indicate that the liquid metal can penetrate the sponges; triangles show that the liquid metal can't penetrate the sponges. The green regions indicate that the liquid metal can penetrate all the tested sponges, and the red region indicates that the liquid metal can't penetrate any of the tested sponges. (e) Diffusion speeds of the penetration effect for sponges A, B, and C under different solution concentrations and applied voltages.

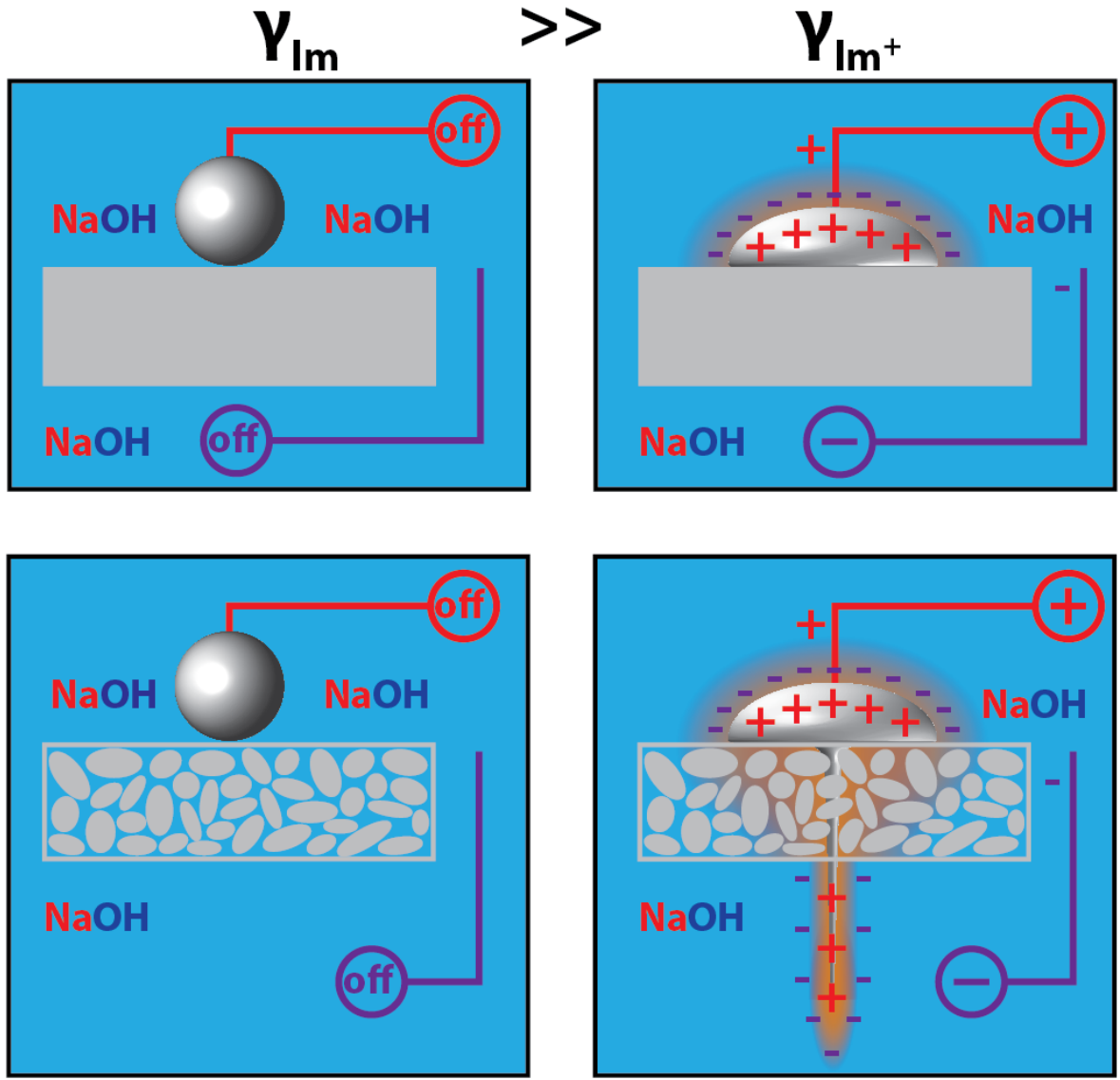

**Fig. 4 | Schematic diagram** of liquid metal spreading and penetration effect for a solid (a, b) and a porous material (c, d) with or without an applied voltage.

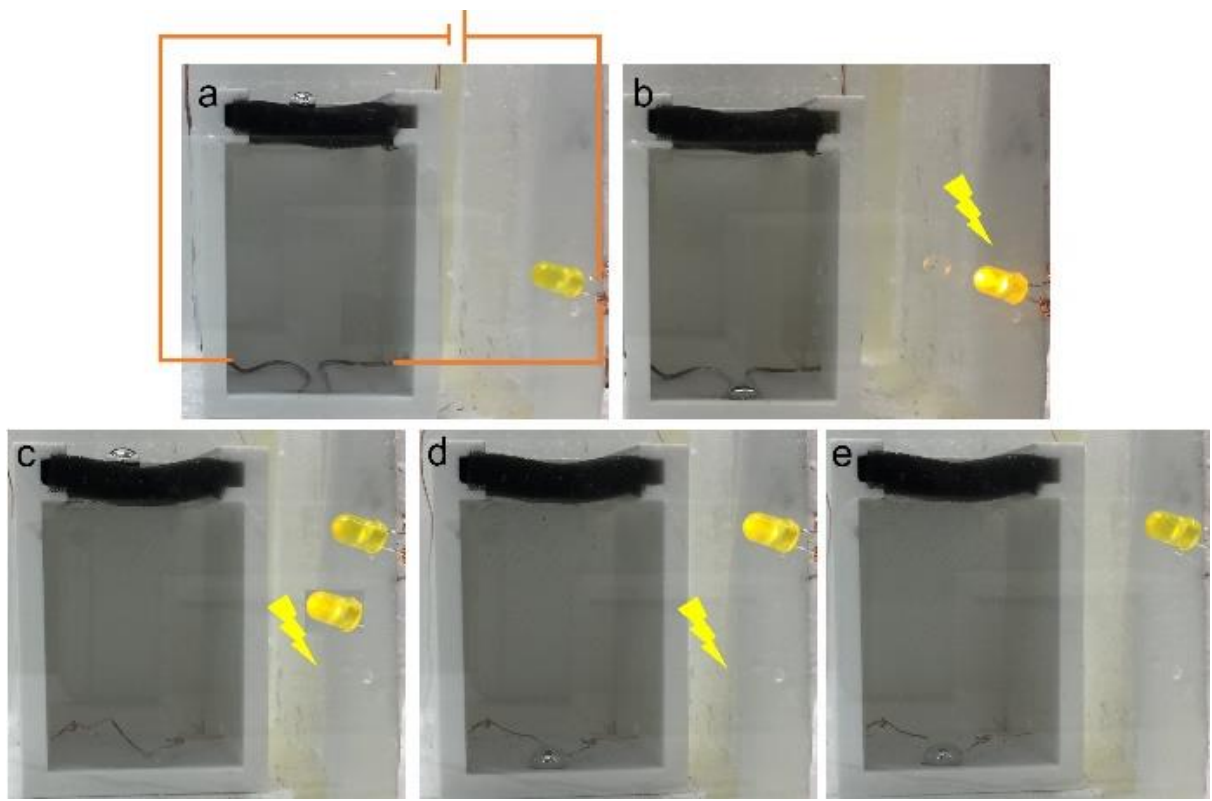

**Fig. 5 | Potential applications**: (a, b) Healing a disconnected circuit in a sealed environment by penetrating a thick sponge. For clarity, the circuit is drawn in (a) using orange lines. (c-e) Cutting-off a closed circuit in a sealed environment.

**Methods**

Galinstan alloy (62 wt % Ga, 22 wt % In, and 16 wt % Sn) with 99.99% purity was purchased from Alfa Aesar. The concentrations of 0.25 - 1 mol/L NaOH solutions were adjusted with a solid 99% NaOH capsule and deionized water. The deionized water was prepared by Purelab Ultra Elga. Three kinds of sponges with different pore sizes and thicknesses were used. Plastic holders were fabricated by using a 3D printer (Me3D, Australia). Other various porous materials such as plastic mesh, fibre mesh, metal mesh, and wiper paper were studied in our experiments. A container of 1 mol/L NaOH solution was prepared at room temperature. The sponge was fixed in a plastic holder, which was put into the container. After a 150 μL droplet of Galinstan was set on the sponge, two copper wires with a diameter of 0.5 mm were connected to the electrolyte and the Galinstan. The applied DC voltages were supplied from 0 to 20 V by a GW laboratory DC power supply (model GPS-1850). A camera was placed in front of the container and recorded the whole progress of the experiment with 1080p horizontal resolution, at 33 frames-per-second (fps).

**Data availability**

All data are available from the corresponding authors and/or included in the manuscript or Supplementary information.

**Acknowledgements**

This work is partially supported by funding from the Australian Research Council through an ARC Future Fellowship project (FT130100778) and a Discovery project (DP130102956).

**Author contributions**

X. W. conceived the project, designed the experiments and led the discussion of experimental results. F. F. Y., Z. Y., and Y. H., X.W carried out experiments and contributed to discussion and analysis of the mechanism. F. F. Y and Z. Y. recorded and analyzed videos of results. X. W., F. F. Y Z. Y., and Y. H. wrote the manuscript.

**Author information**

**Affiliations:**

[1]Institute for Superconducting and Electronic Materials, Australian Institute for Innovative Materials, University of Wollongong, Wollongong, New South Wales 2500, Australia.

[2]Laboratory of Bioinspired Smart Interfacial Science, Technical Institute of Physics and Chemistry, Chinese Academy of Sciences, Beijing 100190, People's Republic of China.

[3]Faculty of Physics & Electronic Sciences, Hubei University, Wuhan, 430062 P.R. China

[4]ARC Centre of Excellence in Future Low-Energy Electronics Technologies, University of Wollongong, Wollongong, New South Wales 2500, Australia.

**Extended Data figures**

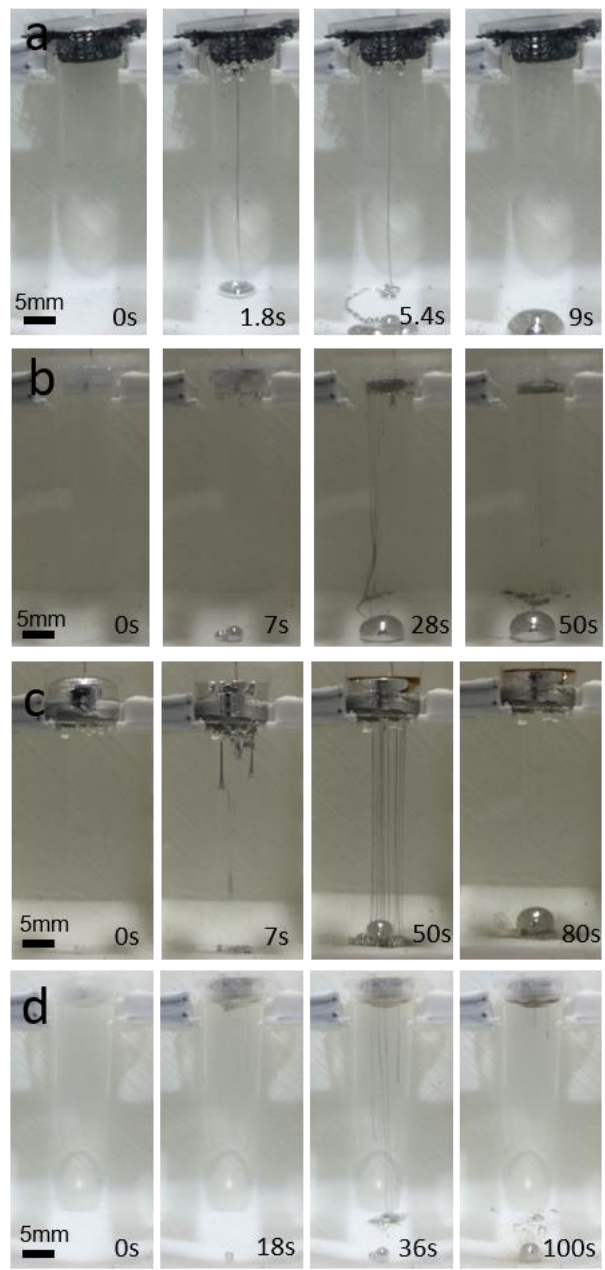

**Extended Data Fig. 1 | Snapshots of the penetration effect** for (a) plastic mesh (pore size of 750 µm), (b) fabric mesh (pore size of ~ 280 mm) (c) metallic mesh (pore size of ~ 45 µm), and (d) wiper paper, with 5 V applied voltage in 1 mol/L NaOH solution.

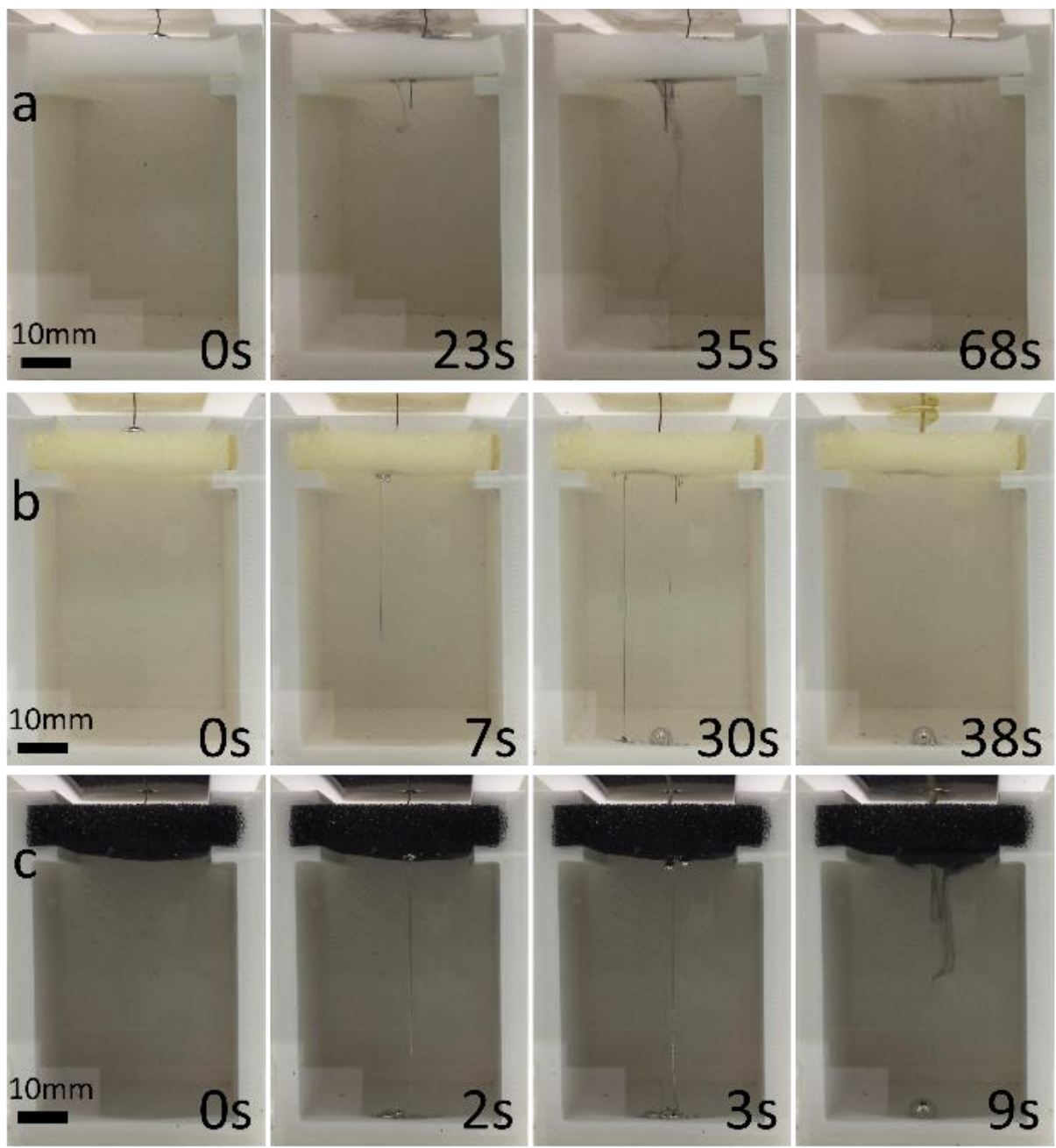

**Extended Data Fig. 2 | Snapshots of the penetration effect** for (a) sponge A (~ 150 mm pore size), (b) sponge B (~ 350 mm pore size), and (c) sponge C (~ 550 mm pore size), each 7.5 mm in thickness with 10 V applied voltage in 1 mol/L NaOH solution.

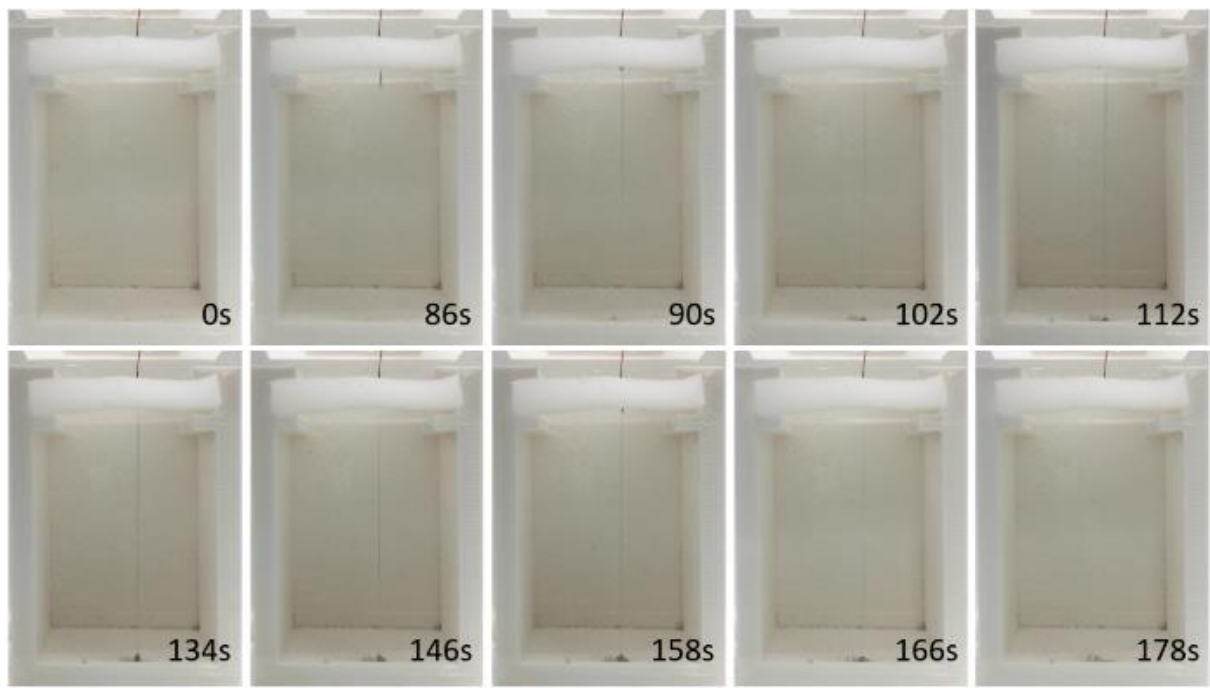

**Extended Data Fig. 3 | Snapshots of the penetration effect** for a sponge A 7.5 mm in thickness with 2.5 V applied voltage in 0.25 mol/L NaOH solution.